\documentclass{article}
\usepackage{spconf,amsmath,graphicx}
\usepackage{bm}
\usepackage{multirow}
\usepackage{array}
\usepackage{cite}
\newcolumntype{M}[1]{>{\centering\arraybackslash}m{#1}}

\title{SOA: Reducing domain mismatch in SSL Pipeline by Speech Only Adaptation for low resource ASR}
\name{Natarajan Balaji Shankar$^1$, Ruchao Fan$^1$, Abeer Alwan$^1$\thanks{This work was supported in part by the NSF.}}
\address{
  $^1$University of California, Los Angeles, Department of Electrical and Computer Engineering\\}
\begin{document}
\ninept
\maketitle
\begin{abstract}

 Recently, speech foundation models have gained popularity due to their superiority in finetuning downstream ASR tasks. However, models finetuned on certain domains, such as LibriSpeech (adult read speech), behave poorly on other domains (child or noisy speech). One solution could be collecting as much labeled and diverse data as possible for joint finetuning on various domains. However, collecting target domain speech-text paired data and retraining the model is often costly and computationally expensive. In this paper, we introduce a simple yet effective method, speech only adaptation (SOA), based on speech foundation models (Wav2vec 2.0), which requires only speech input data from the target domain. Specifically, the Wav2vec 2.0 feature encoder is continually pretrained with the Wav2vec 2.0 loss on both the source and target domain data for domain adaptation, while the contextual encoder is frozen. Compared to a source domain finetuned model with the feature encoder being frozen during training, we find that replacing the frozen feature encoder with the adapted one provides significant WER improvements to the target domain while preserving the performance of the source domain. The effectiveness of SOA is examined on various low resource or domain mismatched ASR settings, including adult-child and clean-noisy speech.
\end{abstract}
\section{Introduction}
\label{sec:intro}

 Self-supervised learning (SSL) has gained popularity in speech processing in recent years \cite{zhang2022bigssl,chen2022wavlm, mohamed2022self, zhang2020pushing, babu2021xls, hsu2021hubert}. SSL is capable of leveraging vast amounts of unannotated data to learn domain knowledge, through a process known as pretraining. This model, containing domain knowledge, can then be used as the starting point for downstream tasks, known as finetuning. SSL models can be utilized in two ways: 1) as a substitute for handcrafted speech features by performing feature extraction \cite{DBLP:conf/interspeech/YangCCLLLLSCLHT21,chang2021exploration}, or 2) as a starting point for model initialization for downstream tasks by performing finetuning \cite{jiang2021further,misra2021comparison,chen2023exploring}.

 One significant drawback of SSL is that training on one domain can lead to domain shifting when finetuning on data from a different domain~\cite{sanabria2022measuring}. Previous studies have attempted to address this issue by incorporating target domain data during pretraining to develop robust pretrained models \cite{hsu2021robust,hwang2022large,wang2022wav2vec,maison2023improving}. In \cite{gururangan2020don}, it is proposed to continually pretrain the base model on in-domain data before commencing finetuning. Previous work such as \cite{khurana2022magic} attempted to use unannotated target domain data for semi-supervised learning for improving ASR performance on unseen languages. In \cite{hsu2021robust}, the effect of adding in-domain data during the pretraining process is analyzed, and the authors note the significant increase in performance and generalizability. The work in \cite{paraskevopoulos2023sample} utilizes a combined loss incorporating labels from the source domain, while simultaneously continually pretraining on the source and target domain.

 Several studies have also attempted to tackle the domain shift problem through the addition of parameters to the base model. In \cite{fan2022draft}, an adapter based approach is proposed to address the domain shift between adult speech data used for pretraining and child speech data used for finetuning. The study in \cite{chen2023chapter} explores the usage of convolutional adapters to the feature encoder for feature adaptation, while  \cite{Bhatia2023} explores the usage of continual pretraining with the presence of adapters for improved ASR performance of accented speech. In \cite{chang2023prompting}, the authors explore the effect of adapter tuning in an encoder-decoder framework for a variety of speech classification and sequence generation tasks.

 While these methods are effective, they either require the presence of extra parameters \cite{fan2022draft, chen2023chapter}, or incur a performance penalty on the original source domain due to finetuning \cite{khurana2022magic, hsu2021robust}. To mitigate this, a proposed solution is to perform joint finetuning utilizing both target domain and source domain data to maintain performance on both sets of data. However, due to the vast size of the training data, this might not always be computationally feasible. The joint finetuning process also necessitates the availability of  transcribed speech-text data from the source domain, as well as the new target low resource domain, both of which might not be readily available.

 In this work, we propose Speech Only Adaptation (SOA), a simple yet effective strategy for increasing performance on the target domain using unlabeled speech data without performance degradation on the source domain. Our adaptation strategy consists of performing continual pretraining on the Wav2vec 2.0 feature encoder using a mix of unlabeled source and target domain data, while keeping the layers of the contextual encoder frozen. We then replace the frozen feature encoder from a source finetuned model with a feature encoder from the continually pretrained model.

The primary advantage of SOA is that it does not require the presence of paired speech-text data from the target domain, and is thus free of any finetuning. Therefore, it can be readily adapted to various low resource scenarios. With models finetuned on the source domain, SOA enables easy adaptation to a target domain with minimal unlabeled target domain data. Furthermore, SOA maintains performance on the source domain while improving performance on the target domain. To validate these claims, we evaluate this technique on two different low resource domain tasks: child (including a zero-shot case) and noisy speech. 

 The remainder of this paper is organized as follows. Section \ref{sec:methods} introduces the framework for the proposed method. Experimental setups are described in Section \ref{sec:exp_setup}. Results are shown and discussed in Section \ref{sec:results}, and we conclude the paper in Section \ref{sec:conclusion}.
\vspace{-10pt}
\section{Methodology}
\label{sec:methods}
\begin{figure}[!h]
\centering{{\includegraphics[width=0.48\textwidth]{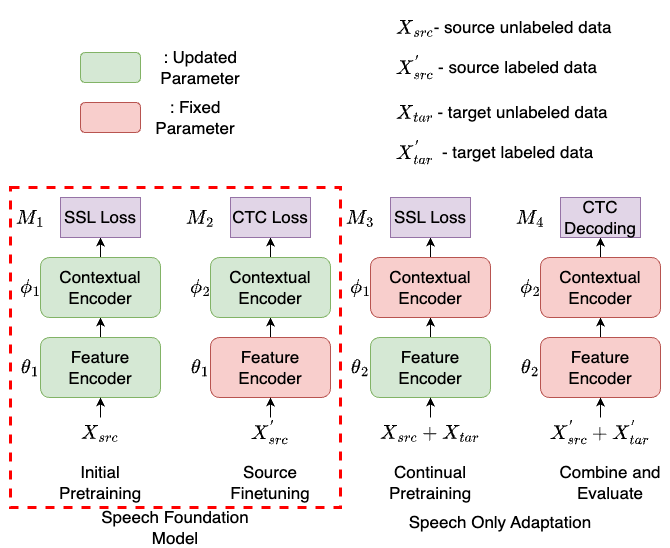} }}%
    \caption{An overview of steps involved in SOA. The steps involved in obtaining the speech foundation model (initial pretraining, and source finetuning) need to be performed just once, and the models obtained can be used for SOA on different target domains}%
    \label{fig:framework}
\end{figure}
\vspace{-10pt}
\subsection{Background}
 Let $x_{src}$ be unlabeled data available from the source domain, $x_{src}',y_{src}$ be paired speech-text (labeled) data from the source domain and $x_{tar}$ be unlabeled data available from the specified target domain. 

 We model the requirements for the pretraining procedure as possessing a large quantity of speech audio data, $x$, typically on the order of hundreds of hours of data. For the purpose of finetuning, we require paired speech-text data, $\{x',y\}$ on the order of tens of hours of data. The pretraining process is modeled as learning a function $f: x \to z$ for learning latent space representations from the raw audio waveforms, and finetuning as learning $f': x\to y$, by iterating on $f$ to obtain the mapping from the audio to the vocabulary for the given dataset. By pretraining and finetuning on the source domain in this manner, we obtain the functions $g$ and $g'$ respectively.

 The issue with a domain shift is that the different distribution of the target domain data from the source domain implies models trained on the source domain do not generalize enough to learn the distribution of the target domain, i.e., $g'$ does not act as a good approximation of $h'$ for mapping between $x_{tar}'$ to $y_{tar}$. The goal of domain adaptation then is to 'adapt' $g'$ to closer approximate $h'$.


 The simplest way to perform this would be to continually train the model on the available $x_{tar}$, thus adapting $g$ to $h$. Even after continual training, the model obtained is a mapping between $x_{tar}$ and $z_{tar}$, and thus without training on a CTC loss cannot be used for ASR. Thus, this requires a way to combine the latter layers of a model finetuned on a CTC loss, which is provided by SOA, as through the combination of encoder model layers, we approximate the iterating process to obtain $h'$.

\subsection{SSL Pipeline - Pretraining and Finetuning}
 In this work, we utilize the base Wav2vec 2.0 model \cite{baevski2020wav2vec} as the speech foundation model for its relative computational simplicity. The model consists of a convolutional feature encoder $\theta  : x \to z$ to map raw audio x to latent representations $z_1 , \dots, z_N$. These representations are input to a transformer based contextual encoder $\phi : z \to c$ to output context representations $c_1, \dots, c_N$. During pretraining, latent representations are discretized to $q_1,\dots, q_N$  using a vector quantization module. The model is trained to identify the true quantized latent $q_t$ using $c_t$ for each masked time step within a set of distractors sampled from other masked time steps using a contrastive loss \cite{baevski2020wav2vec}. During finetuning, we shift to using a Connectionist Temporal Classification (CTC) loss \cite{graves2006connectionist} objective, and task the model with predicting the characters of the paired text data using the representations $c_t$.

 Let the speech foundation model (Wav2vec 2.0) be represented by $ M_1 : \{\theta_1, \phi_1\}$, where $\theta_1$ represents the parameters present in the feature encoder and $\phi_1$ represents the parameters of the contextual encoder. Finetuning of $M_1$ using paired data $\{x_{src}', y_{src}\}$, while keeping the weights in the feature encoder frozen, results in model $M_2: \{\theta_1, \phi_2 \}$, where $M_2$ is now a model optimized for performance on  the source domain. This model $M_2$ can be reused for multiple target domains.
 
\subsection{Proposed Framework} 

 SOA is a two-stage (continual pretraining, and combination) training paradigm as shown in Figure \ref{fig:framework}.  
  SOA can be described as:
 
\textbf{Stage 1:} Continual pretraining of $M_1$ using the Wav2vec 2.0 loss  with data $x_{src} \cup x_{tar}$ while keeping the parameters of $\phi_1$ frozen. This results in model $M_3: \{\theta_2, \phi_1 \}$. 

\textbf{Stage 2:} Combining the contextual encoder $\phi_2$ from $M_2 $ with the feature encoder $\theta_2$ from $M_3$, resulting in $M_4: \{ \theta_2, \phi_2 \}$. 

 Note that by freezing the parameters of $\theta_1$ during finetuning on the source data, and $\phi_1$ during continual pretraining, catastrophic forgetting is effectively prevented.

 The advantage in the usage of SOA is twofold: 1) As only the contrastive loss during pretraining is employed for the target domain adaptation, we do not require any paired speech-text data $\{x_{tar}', y_{tar} \}$, which is difficult to procure for low resource tasks. 2) Models $M_1$ and $M_2$ pretrained and finetuned on the source domain can be reused for subsequent adaptation to several target domains, thus reducing the computational cost.

\section{Experimental Setup}
\label{sec:exp_setup}
\subsection{Datasets}
\label{ssec:datasets}

 Our experiments utilized a diverse array of datasets, each categorized as follows. The foundational base models were initially pretrained using the LibriSpeech corpus \cite{PanayotovCPK15}, consisting of 960 hours of read adult speech. We refer to the model optimized for clean, adult speech as $M_2$, achieved through finetuning on the 100-hour clean section of this dataset. In pursuit of enhancing performance for various low resource tasks, we utilized the following datasets for the continuous pretraining of $M_3$.

\subsubsection{Children's Speech}

 For child speech experimentation, we employed the MyST Children's Speech Corpus \cite{ward2011my}, encompassing a total of 499 hours of speech data comprising 244,069 conversational utterances exchanged between children and a virtual tutor. This dataset involved interactions from 1,372 students spanning the third to fifth grades. However, only 42\% of the corpus, equivalent to 240 hours, has ASR annotations. The corpus also contains dedicated development and test sets designed for evaluation purposes. We investigate training on varying subsets of this data to assess the efficacy of our framework. 

 In addition, we conducted assessments of our method's performance using the CMU Kids Corpus \cite{eskenazi1997cmu} to illustrate transferability of performance across different children's speech corpora.  The corpus contains 5180 utterances of read speech from 76 speakers, amounting to a total of 9 hours of child speech. We utilize the entirety of the corpus to perform zero-shot inference to demonstrate the efficacy of SOA.



\subsubsection{Noise Robustness}
 To evaluate the performance of our framework on noisy speech, we generate noisy speech files through the addition of noise obtained from the MUSAN dataset \cite{snyder2015musan} to files from the clean section of the LibriSpeech dataset. To simulate noisy speech data, we randomly select noise samples from FreeSound \cite{font2013freesound} subset of the MUSAN corpus and mix them with the LibriSpeech clean samples, at a randomly selected SNR from [0, 15] dB. We evaluate the performance of this method through the creation of noisy dev and test sets by a similar process of addition of noise to the dev-clean and test-clean sections of the LibriSpeech set as well.


\subsection{Model Settings}
 For the base foundation model $M_1$, we use the open-sourced base Wav2vec 2.0 model (95M parameters) from the fairseq toolkit \cite{ott2019fairseq}. During different stages of the framework, we obtain different models with varying numbers of updated parameters. Note that $M_2$ (with updated encoder layers) has close to 92M updated parameters, while $M_3$ (with only updated convolution layer weights) has close to 5M updated weights, thus leading to an efficient finetuning process for domain adaptation, as $M_2$ does not need to be retrained for every new domain task.

 For the finetuning of $M_2$, we update using a noam scheduler \cite{vaswani2017attention} with warm up steps of 8k, and a peak learning rate of 3e-5. The peak learning rate holds for the next 32k steps, then exponentially decays to the ratio $\lambda$ of the initial learning rate, where $\lambda$ is set to 0.05.

 We perform continual pretraining of $M_3$ using the Adam optimizer. During the first 8\% of all training steps, the learning rate increases to 3e-5 and then decays polynomially. We experiment with different target domains $x_{tar}$, as well as varying the size of source $x_{src}$ and target $x_{tar}$ domain corpus used for SOA.

 Training for all models is conducted on 2 Nvidia A4000 GPUs. The number of Floating Point Operations needed for training is computed by multiplying the training time of the model, the number of GPUs used during training, and an estimate of the single-precision floating-point capacity of the GPU (19.17 TFLOPS).

\section{Results and Discussion}
\label{sec:results}
\subsection{Child Speech}
\subsubsection{Baseline}
 We first offer a comparison of our method (SOA) with the performance of other similarly sized models in Table \ref{tab:baseline}. For a fair comparison, we list the respective Word Error Rates of HuBERT \cite{hsu2021hubert}base models finetuned on LibriSpeech. We use a 4-gram LM trained on the LibriSpeech corpus for decoding. The performance on these corpora is similar to previously published results \cite{fan2022draft, fan2022towards}. Our SOA model listed involved unsupervised domain adaptation of model $M_3$ till convergence (100k updates) using the 240 hour MyST train corpus with the 100 hour LibriSpeech train corpus. 

 We also offer a comparison of our method (SOA) with other domain adaptation methods in Table \ref{tab:table2}, as well as an estimate of the time required for the adaptation. The jointly finetuned model can be viewed as an upper bound on the performance on the specified task. Finetuning on just MyST leads to reduced performance on the source domain compared to both the LibriSpeech finetuned model and SOA. We also continually pretrain a model on the target domain data before finetuning and utilize the M2DS2 framework \cite{paraskevopoulos2023sample}.  Note that SOA outperforms other unsupervised methods while requiring less computational effort, highlighting its effectiveness.

 We note that across the Wav2vec 2.0 and HuBERT models that finetuning on either the LibriSpeech or MyST corpus leads to a loss of generalizability and subsequent performance on the other corpus, and the SOA model is able to maintain performance on the LibriSpeech (source) domain, while reducing the WER on the MyST (target) domain. SOA also does not require the presence of any labeled data from the MyST dataset, while the finetuned model utilizes 240 hours of labeled data for its improved WER. We note that the SOA model is able to achieve a 6.44\% reduction in relative Word Error Rate, a statistically significant ($p < 0.05$) result compared to the baseline performance.
\begin{table}[!htbp]\centering
\resizebox{\columnwidth}{!}{%
\scriptsize
\begin{tabular}{|c|c|c|c|c|c|c|c|}
\hline
\multirow{2}{*}{Model} &\multirow{2}{*}{Decoding} &\multicolumn{4}{c|}{LibriSpeech} &\multicolumn{2}{c|}{MyST} \\[5pt]\cline{3-8}
& &dev-clean &dev-other &test-clean &test-other &development &test 
\\[5pt]
\hline
HuBERT &
\begin{tabular}[c]{@{}c@{}}w/o LM\\ LM\end{tabular} & 
\begin{tabular}[c]{@{}c@{}}5.31\\ 3.19\end{tabular} &
\begin{tabular}[c]{@{}c@{}}13.15\\ 8.45\end{tabular} &
\begin{tabular}[c]{@{}c@{}}5.39\\ 3.44\end{tabular} &
\begin{tabular}[c]{@{}c@{}}12.77\\ 8.32\end{tabular} &
\begin{tabular}[c]{@{}c@{}}32.82\\ 26.26\end{tabular} &
\begin{tabular}[c]{@{}c@{}}35.15\\ 28.62\end{tabular} \\[8pt]
\hline
Wav2vec 2.0 &
\begin{tabular}[c]{@{}c@{}}w/o LM\\ LM\end{tabular} & 
\begin{tabular}[c]{@{}c@{}}5.33\\ 3.12\end{tabular} &
\begin{tabular}[c]{@{}c@{}}13.85\\ 8.76\end{tabular} &
\begin{tabular}[c]{@{}c@{}}5.41\\ 3.39\end{tabular} &
\begin{tabular}[c]{@{}c@{}}13.15\\ 8.65\end{tabular} &
\begin{tabular}[c]{@{}c@{}}32.72\\ 26.45\end{tabular} &
\begin{tabular}[c]{@{}c@{}}35.07\\ 28.76\end{tabular} \\[8pt]
\hline
SOA &
\begin{tabular}[c]{@{}c@{}}w/o LM\\ LM\end{tabular} & 
\begin{tabular}[c]{@{}c@{}}5.34\\ 3.21\end{tabular} &
\begin{tabular}[c]{@{}c@{}}14.19\\ 8.94\end{tabular} &
\begin{tabular}[c]{@{}c@{}}5.48\\ 3.43\end{tabular} &
\begin{tabular}[c]{@{}c@{}}13.47\\ 8.84\end{tabular} &
\begin{tabular}[c]{@{}c@{}}\textbf{30.59}\\ \textbf{24.91}\end{tabular} &
\begin{tabular}[c]{@{}c@{}}\textbf{32.81}\\ \textbf{27.13}\end{tabular} \\[8pt]
\hline
\end{tabular}
}
\caption{WER results of HuBERT, Wav2vec 2.0 and SOA on the LibriSpeech and MyST datasets. The HuBERT and Wav2vec 2.0 models were finetuned on the LibriSpeech dataset. A 4-gram LibriSpeech model was used for LM decoding}\label{tab:baseline}
\end{table}

\begin{table}[!htbp]\centering
\resizebox{\columnwidth}{!}{%
\scriptsize
\begin{tabular}{|c|c|c|c|c|c|c|}
\hline
\multirow{2}{*}{Model} &\multicolumn{2}{c|}{LibriSpeech} &\multicolumn{2}{c|}{MyST} & Training Cost \\\cline{2-5}
  &test-clean &test-other &development &test & (FLOPs)
\\
\hline
LibriSpeech finetuned &  5.41 & 13.15 & 32.72 & 35.07 &-\\
\hline
\multicolumn{6}{|c|}{Supervised Methods} \\ \hline

MyST finetuned & 22.38	&35.22	&13.45	&15.08 &-\\
\hline
Jointly finetuned &  6.08 & 14.03 & 14.08 & 15.57 &-\\
\hline
\multicolumn{6}{|c|}{Unsupervised Methods} \\ \hline
Continual Pretraining & 5.45 & 13.41 & 32.05 & 34.2 & $6.8 \cdot 10^{18}$\\
\hline
M2DS2 \cite{paraskevopoulos2023sample}& 5.70 &  13.19 & 32.43 & 34.46 & $2.6 \cdot 10^{18}$\\
\hline
SOA (Ours)  & 5.48 & 13.47 & \textbf{30.59} & \textbf{32.81} & $\bm{1.5 \cdot 10^{18}}$\\
\hline
\end{tabular}
}
\caption{WER results of different domain adaptation methods on the LibriSpeech and MyST datasets. The training cost in Floating Point Operations (FLOPs) is estimated as detailed in Section 3.2}\label{tab:table2}
\end{table}

\subsubsection{Effect of the size of Low Resource Domain Data}
 To evaluate the effect of the size of the pretraining corpus used for SOA, we repeat the unsupervised domain adaptation experiments by varying the amount of target domain data $x_{tar}$ available. For all of these experiments, we keep the amount of source domain data $x_{src}$ fixed by utilizing the entire LibriSpeech corpus (100 hours). In Table \ref{tab:target_domain_hours} we present the non-LM decoding WERs of these models.  We note that, as expected, an increase in the size of the target domain corpus available for SOA leads to a reduction in the WER of the model, with an effect noted even using just 1 hour of data for SOA.
\begin{table}[!htbp]\centering
\resizebox{\columnwidth}{!}{%
\begin{tabular}{|c|c|c|c|c|c|c|}
\hline
MyST&\multicolumn{4}{c|}{LibriSpeech} &\multicolumn{2}{c|}{MyST} \\\cline{2-7} Training Data &dev-clean &dev-other &test-clean &test-other &development &test \\
\hline
Baseline & 5.33	&13.85	&5.41	&13.15	&32.72	&35.07 \\
\hline
1h &5.33 &14.04 &5.42 &13.25 &32.52 &34.72 \\
\hline

10h &5.35 &14.00 &5.48 &13.29 &31.85 &34.09 \\
\hline

100h &5.36 &14.16 &5.49 &13.40 &30.92 &33.14 \\
\hline
240h &5.34 &14.10 &5.48 &13.48 &\textbf{30.66} &\textbf{32.90} \\
\hline
\end{tabular}%
}
\caption{WER results of SOA models trained using a different number of hours of the MyST training data on the LibriSpeech and MyST datasets}\label{tab:target_domain_hours}
\end{table}
\subsubsection{Zero-shot performance on auxiliary children's corpus}
 Thus far, we have evaluated the performance of SOA with the target domain data $x_{tar}$ obtained from the MyST corpus through testing on the same corpus. To demonstrate the effectiveness of the method in learning features inherent to children's speech, we test the zero-shot performance of SOA performed using varying amounts of the MyST corpus on the CMU Kids dataset, as shown in Table \ref{tab:cmu_data}. For these experiments, we do not use any amount of the training corpus from the CMU Kids corpus for either unsupervised domain adaptation or for finetuning. We note that an increase in the amount of MyST data used in the SOA process leads to a decrease in WER on the CMU Kids corpus as well, with training on the entire MyST corpus resulting in an 8.06\% reduction in relative Word Error Rate.
\begin{table}[!htbp]\centering

\resizebox{0.7\columnwidth}{!}{%
\begin{tabular}{|*{4}{c|}}
\hline
MyST&\multicolumn{2}{|c|}{MyST} &CMU\\\cline{2-3}
Training Data  &development &test & Kids\\
\hline
Baseline	&32.72 &	35.07 & 35.35\\
\hline
1h 	&32.52	&34.72 &	35.27\\
\hline
10h 	&31.85	&34.09 & 33.93\\
\hline
100h 	&30.92 &	33.14 & 32.91\\
\hline
240h 	&30.66&	32.90  &\textbf{32.50}\\
\hline
\end{tabular}%
}
\caption{ Zero-shot WER results of SOA models trained using a different number of hours of the MyST training data on the MyST and CMU Kids datasets}\label{tab:cmu_data}
\end{table}

\vspace{-5pt}

\subsection{Noise Robustness}
 We also evaluate the effectiveness of SOA on adaptation for noisy speech in Table \ref{tab:noisy_data} through the variation in the number of training hours of the noisy (target) data used for SOA. Here, we refer to Baseline as model $M_2$ finetuned on only LibriSpeech (source) data without any SOA, and note that the performance for the baseline method is consistent with previously published results \cite{hu2023wav2code}. SOA was performed for all the listed models for a total of 50k updates. We note that by utilizing 100 hours of the target domain, we are able to achieve a 28.9\% relative Word Error Rate reduction on the noisy test set. We also report the results from decoding using 4-gram LibriSpeech LM for the baseline and best performing SOA model.
\begin{table}[!htbp]\centering
\resizebox{\columnwidth}{!}{%
\begin{tabular}{|c c|c|c|c|c|c|c|}
\hline
\multicolumn{2}{|c|}{Noisy}&\multicolumn{4}{c|}{LibriSpeech} &\multicolumn{2}{c|}{Noisy} \\[5pt]\cline{3-8} \multicolumn{2}{|c|}{Training Data} &dev-clean &dev-other &test-clean &test-other &development &test \\[5pt]
\hline
 Baseline& \begin{tabular}[c]{@{}c@{}}w/o LM\\ LM\end{tabular} & 
\begin{tabular}[c]{@{}c@{}}5.33\\ 3.12\end{tabular} &
\begin{tabular}[c]{@{}c@{}}13.85\\ 8.76\end{tabular} &
\begin{tabular}[c]{@{}c@{}}5.41\\ 3.39\end{tabular} &
\begin{tabular}[c]{@{}c@{}}13.15\\ 8.65\end{tabular} &
\begin{tabular}[c]{@{}c@{}}16.87\\12.54\end{tabular} &
\begin{tabular}[c]{@{}c@{}}15.29\\11.2\end{tabular} \\[8pt]
\hline
\multicolumn{2}{|c|}{1h} & 5.33	& 13.98 & 	5.44	& 13.25 &	16.03	& 14.46 \\[3pt]
\hline

\multicolumn{2}{|c|}{10h}& 5.33	&13.92	&5.41	&13.24	&13.83	&12.52 \\[3pt]
\hline

100h & \begin{tabular}[c]{@{}c@{}}w/o LM\\ LM\end{tabular} & 
\begin{tabular}[c]{@{}c@{}}5.33\\ 3.15\end{tabular} &
\begin{tabular}[c]{@{}c@{}}13.91\\ 8.79\end{tabular} &
\begin{tabular}[c]{@{}c@{}}5.44\\ 3.38\end{tabular} &
\begin{tabular}[c]{@{}c@{}}13.28\\ 8.75\end{tabular} &
\begin{tabular}[c]{@{}c@{}}\textbf{12.11}\\\textbf{8.35}\end{tabular} &
\begin{tabular}[c]{@{}c@{}}\textbf{10.86}\\\textbf{7.64}\end{tabular} \\[8pt]
\hline
\end{tabular}%
}
\caption{WER results of SOA models trained using a different number of hours of the noisy training data on the LibriSpeech and Noisy datasets}\label{tab:noisy_data}
\end{table}
\vspace{-5pt}
\subsection{Feature Encoder Output Analysis}
\vspace{-5pt}
\begin{figure}[!htbp]
\centering{{\includegraphics[width=0.35\textwidth]{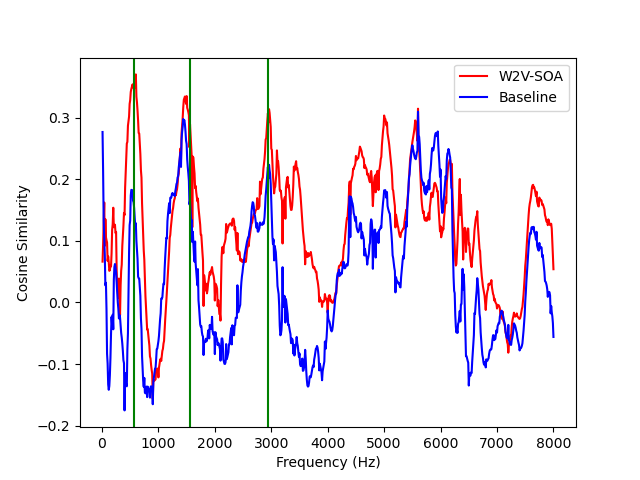} }%
    \caption{ Cosine Similarity between feature space representations of pure sinusoids and sample audio for baseline model (blue) and SOA model
(red). The formant frequencies for the audio file lie at 568 Hz, 1559 Hz and 2944 Hz respectively (green).}\label{fig:u_plot}
}%

\end{figure}
To explore how the SOA model shifts the representations of the features vectors $z$, we perform the following experiment. We first feed both the baseline and SOA models a 1-second signal $ x_1 = \sin(2\pi f_l t)$ with $f_l$ ranging from
10Hz to 8kHz at 10Hz intervals, as in \cite{choi2022opening}, to obtain the output representation $z_1$. We then feed the models with a speech signal $x_2$ from the MyST dataset, whose output representation is $z_2$. We proceed to compute the cosine similarity between $z_1$ and $z_2$ for different values of $f_l$ to demonstrate the ability of the feature encoder to 'capture' the formant frequency.

Figure \ref{fig:u_plot} demonstrates a plot of the cosine similarities of the feature space encodings as a function of frequency for the vowel in the word 'Good' from the MyST dataset. We note that there is a spike in the values of the cosine similarities at the formant frequencies for the vowel ($F_1:568 $Hz$, F_2: 1559 $Hz$, F_3: 2944 $Hz). We see that the SOA model shows sharper and more pronounced peaks, indicating that the SOA method leads to the feature encoder being more attuned to the formant frequencies of the target domain. While this plot is only for an individual audio sample, this trend holds up across a variety of utterances, indicating that this shift in the representation space could be the reason behind the improved performance of SOA.

\section{Conclusion}
\label{sec:conclusion}

In this paper, we introduce a novel method, Speech Only Adaptation (SOA), for utilizing unlabeled data from a target domain to perform unsupervised domain adaptation. Specifically, by continually pretraining the feature encoder while keeping the contextual encoder frozen, and replacing the frozen contextual encoder with one obtained during finetuning we demonstrate that it is possible to improve performance on a low resource target domain, while maintaining performance on the source domain. When compared to the conventional finetuning
baselines without adaptation, we achieved relative WER improvements of up to 6.4\% on the MyST child ASR, and 28.9\% on noisy ASR, demonstrating the efficacy of this method. We also illustrate the cross-corpus transferability of performance through an 8.06\% relative WER reduction on zero-shot evaluation of the CMU Kids corpus. In scenarios where one can only access unlabeled data (e.g., YouTube recordings), either directly from the target domain or from a closely-related distribution, 
SOA allows the reuse of finetuned source domain models, making the proposed framework promising for future low resource ASR tasks.


\footnotesize
\bibliographystyle{IEEEbib}
\bibliography{strings,refs}

\end{document}